\documentclass[
    ,final            
  ]
{aipproc1}

\layoutstyle{8x11single}

\begin{document}

\title{CMS results on multijet correlations}

            

\author{Grigory Safronov \thanks{on behalf of the CMS collaboration}}{
  address={ITEP, B. Cheryomushkinskaya, 25, 117218, Moscow, Russia},
  email={safronov@itep.ru}
}

\begin{abstract}
 We present recent measurements of multijet correlations using forward and low-$p_{\mathrm{T}}$ jets performed by the CMS collaboration at the LHC collider. In pp collisions at $\sqrt{s} = 7$ TeV,  azimuthal correlations in dijets separated in rapidity by up to 9.4 units were measured. The results are compared to BFKL- and DGLAP-based Monte Carlo generator and analytic predictions. In pp collisions at $\sqrt{s} = 8$ TeV, cross sections for jets with $p_{\mathrm{T}}$ > 21 GeV and |y| < 4.7, and for track-jets with $p_{\mathrm{T}}$ > 1 GeV (minijets) are presented. The minijet results are sensitive to the bound imposed by the total inelastic cross section, and are compared to various models for taming the growth of the $2 \rightarrow 2$ cross section at low $p_{\mathrm{T}}$.\end{abstract}

\maketitle


\section{Introduction}

Hard parton-parton interactions occurring during hadron collisions can be described in the theory of Quantum chromodynamics (QCD).
Hadronic jets, collimated streams of particles carrying information about partons produced in the hard interaction, are particularly useful probes for studies of QCD. Perturbative QCD (pQCD) calculations at next-to-leading (NLO) order accuracy, using the collinear factorisation framework and Dokshitzer-Gribov-Lipatov-Altarelli-Parisi (DGLAP) equations are well-tested tools describing inclusive jet or dijet production. Measurements of correlations of jets within the same event are sensitive to higher order pQCD radiation (or parton showers), which is taken into account in analytic calculations by the resummation procedure. The DGLAP equations account for such radiation to all orders of perturbation theory for hard parton collisions ($\sqrt{s} \sim p_{\mathrm{T}} \gg \Lambda_{QCD}$). At high centre-of-mass energies a kinematical domain can be reached where semi-hard parton interactions ($\sqrt{s} \gg p_{\mathrm{T}} \gg \Lambda_{QCD}$) play a substantial role. Perturbative QCD resummation in the asymptotic region, where $\sqrt{s} \to \infty $, is performed by the Balitsky-Fadin-Kuraev-Lipatov (BFKL) equation. 

In this report we review several CMS measurements probing a number of QCD properties. Measurement of the low-$p_{\mathrm{T}}$ jet integrated cross-section \cite{minij} probes transition from perturbative non-perturbative QCD domain. Low-$p_{\mathrm{T}}$ jet spectrum measurement \cite{lowpt} allows testing of fixed order pQCD calculations. Finally correlations between jets widely separated in rapidity \cite{Chatrchyan:2012pb, azimu} probes phase space region where pQCD resummation techniques are particularly important for data description. Therefore it provides testing ground for different parton shower models and analytic calculations.

\section{The CMS detector}

The calorimeter system of the CMS detector \cite{cmsexp} covers pseudo-rapidity range $\vert \eta \vert < 5.0$, where $\eta=-\log[\tan(\theta/2)]$, and $\theta$ is the polar angle relative to the anticlockwise proton beam direction. The crystal electromagnetic calorimeter (ECAL) and the brass/scintillator hadronic calorimeter (HCAL) extend to pseudorapidities $\vert \eta \vert = 3.0$. The HCAL cells map to an array of ECAL crystals to form calorimeter towers projecting radially outwards from 
the nominal interaction point. The pseudorapidity region $3.0<\vert \eta \vert <5.0$ is covered by the hadronic forward (HF) calorimeter.
Tracking detector which consists of silicon pixel vertexing detector surrounding the interaction point and silicon strip tracker covers pseudo rapidity range $\vert \eta \vert < 2.4$.

Several methods of jet reconstruction are used by the CMS collaboration. Calorimeter jets are clustered from energy deposits in calorimeter towers. Particle flow jets are formed by clustering particles as reconstructed by the particle flow algorithm using information from all subdetectors. The anti-$k_{\mathrm{T}}$ algorithm is used for jet clustering. The precision of the jet energy calibration varies for different jet types \cite{JME-10-011pub}, and is within~$7-8\%$ for forward jets (at $p_{\mathrm{T}} \simeq 35$ GeV and $3 < \vert \eta \vert < 4.7$).

\section{Single jet production cross section measurements}

\begin{figure}[hbtp]
    \includegraphics[width=0.4 \textwidth]{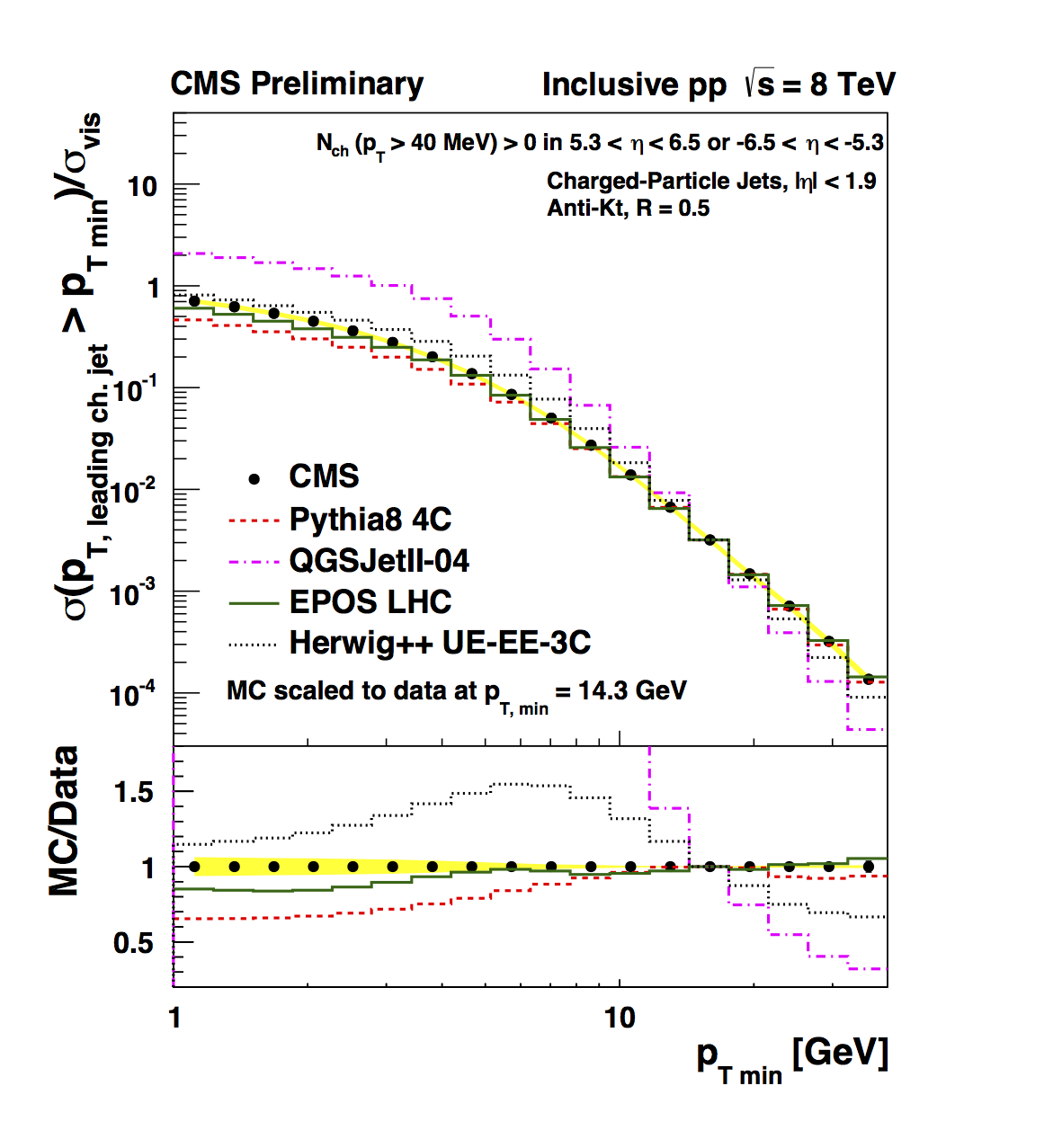} 
    \includegraphics[width=0.4\textwidth]{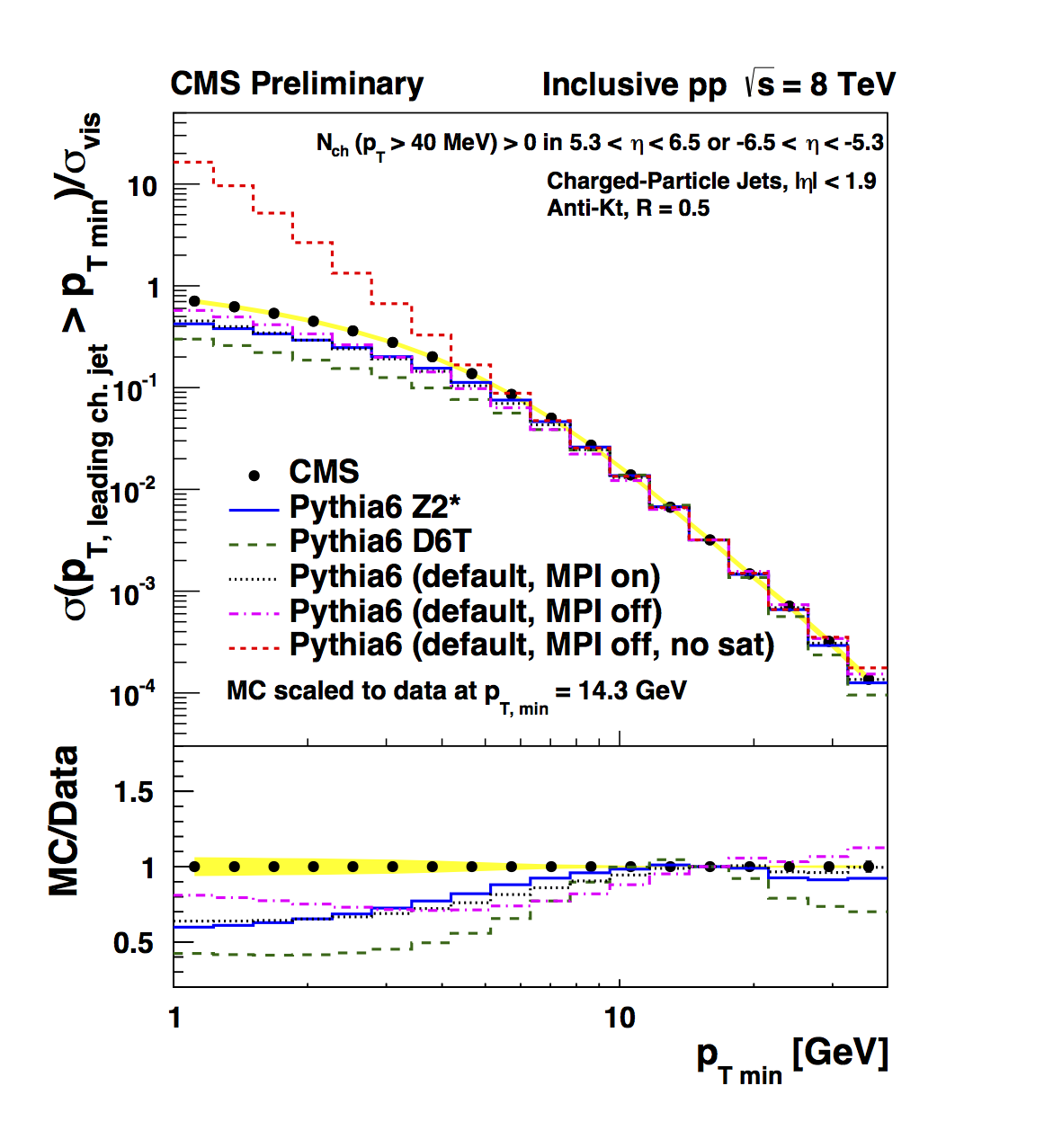} 
     \caption{Normalised integrated cross section for minijet production as a function of $p_{\mathrm{T,min}}$. Measured cross-section is compared to various Monte Carlo predictions. Monte Carlo predictions are normalised to match the measured normalised cross section value for $p_{\mathrm{T,min}}=14.3$ GeV.}
    \label{fig:minijets}
\end{figure}

Most of the final state hadrons produced in proton-proton interactions at LHC energies come from "semi-hard" interactions where exchange momenta is small ($O(1 - 3)$GeV). At low $p_{\mathrm{T}}$ the cross-section for hadron production is becoming very large and eventually total integrated cross-section $\sigma(p_{\mathrm{T,min}}) = {\int_{p_{\mathrm{T,min}}}} dp_{\mathrm{T}}^{2} d \sigma /p_{\mathrm{T}}^4$ overcomes the total proton-proton inelastic cross-section. In \cite{Grebenyuk:2012qp} it was proposed to measure the small-$p_{\mathrm{T}}$ jet cross-section integrated over the $p_{\mathrm{T,min}}$ as a probe for the transition from perturbative to non-perturbative QCD domain. Measurement \cite{minij} was performed with data recorded at collision energy of 8 TeV in common data-taking with TOTEM experiment \cite{Anelli:2008zza} in a dedicated LHC run with very low number of overlapping collisions ($\sim 5.4\%$ on average). T2 telescope of TOTEM detector provides tracking in $5.3 < \vert \eta \vert < 6.5$ region. Events were selected with trigger requiring at least 1 track with $p_{\mathrm{T}}$ > 40 MeV in T2 acceptance. Analysed data corresponds to 45 $\mu$b$^{-1}$. Jets with $p_{\mathrm{T}}>1$ GeV and $\vert \eta \vert < 1.9$ clustered from charged hadrons reconstructed with tracking detector were selected for the analysis. Integrated jet cross-section was measured in the range $1 < p_{\mathrm{T,min}} < 40$ (GeV). Measured cross section was normalised to total cross-section of production of jets satisfying selection criteria. Results of the measurement are presented at Fig.~\ref{fig:minijets}. Taming of the cross section for low $p_{\mathrm{T,min}}$ is observed. Measured cross section was compared to predictions of various Monte Carlo (MC) models. It was observed that the shape of integrated cross section distribution as obtained from data is well-described by cosmic ray MC \textsc{epos} \cite{Pierog:2013ria}. While well-known MC \textsc{pythia},  \textsc{herwig} and other fail to describe the taming of the cross section. Analogous analysis discussed in \cite{minij} is made for leading track objects. It has similar conclusions.

\begin{figure}[hbtp]
    \includegraphics[width=0.33 \textwidth]{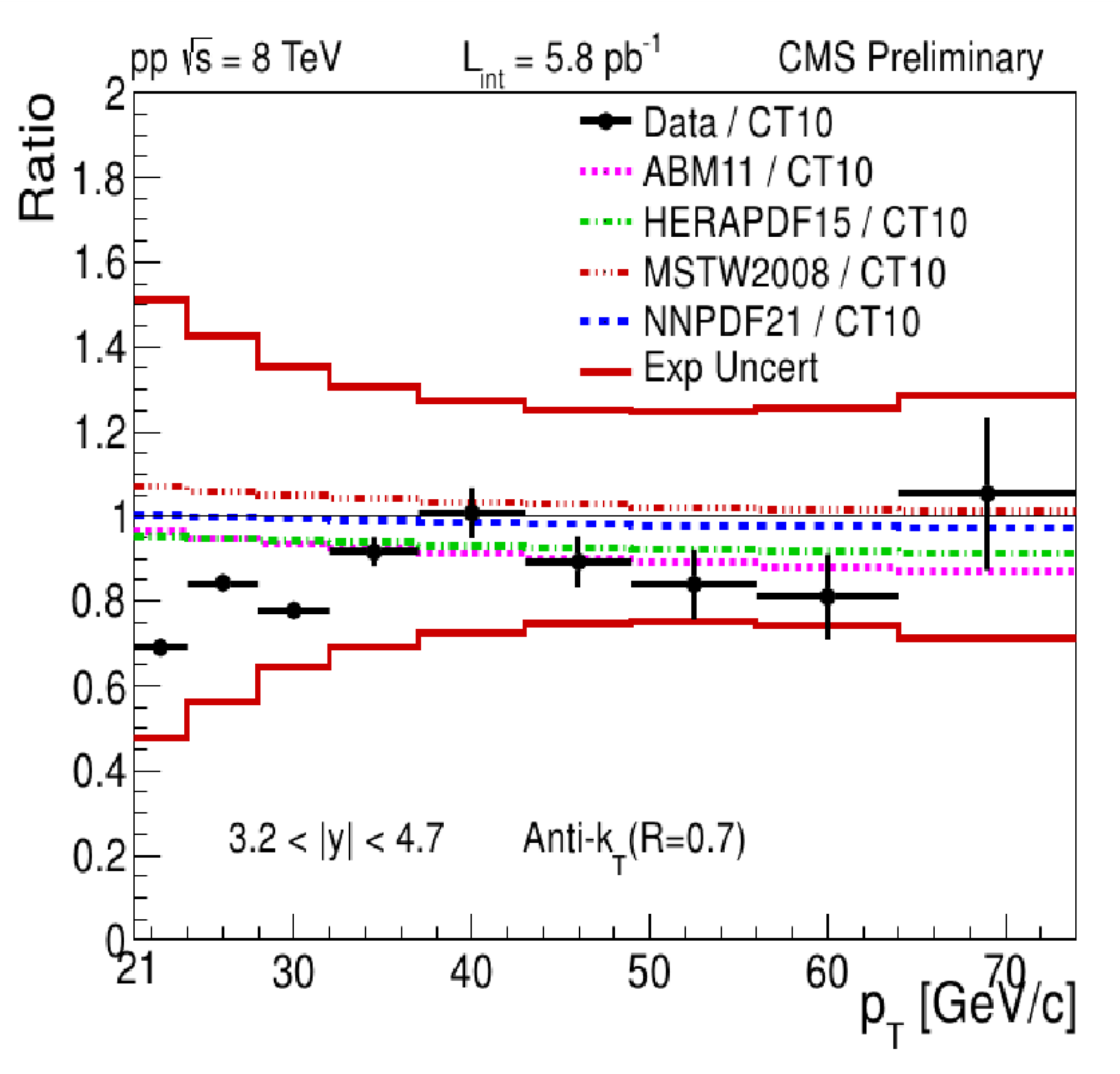} 
    \includegraphics[width=0.4\textwidth]{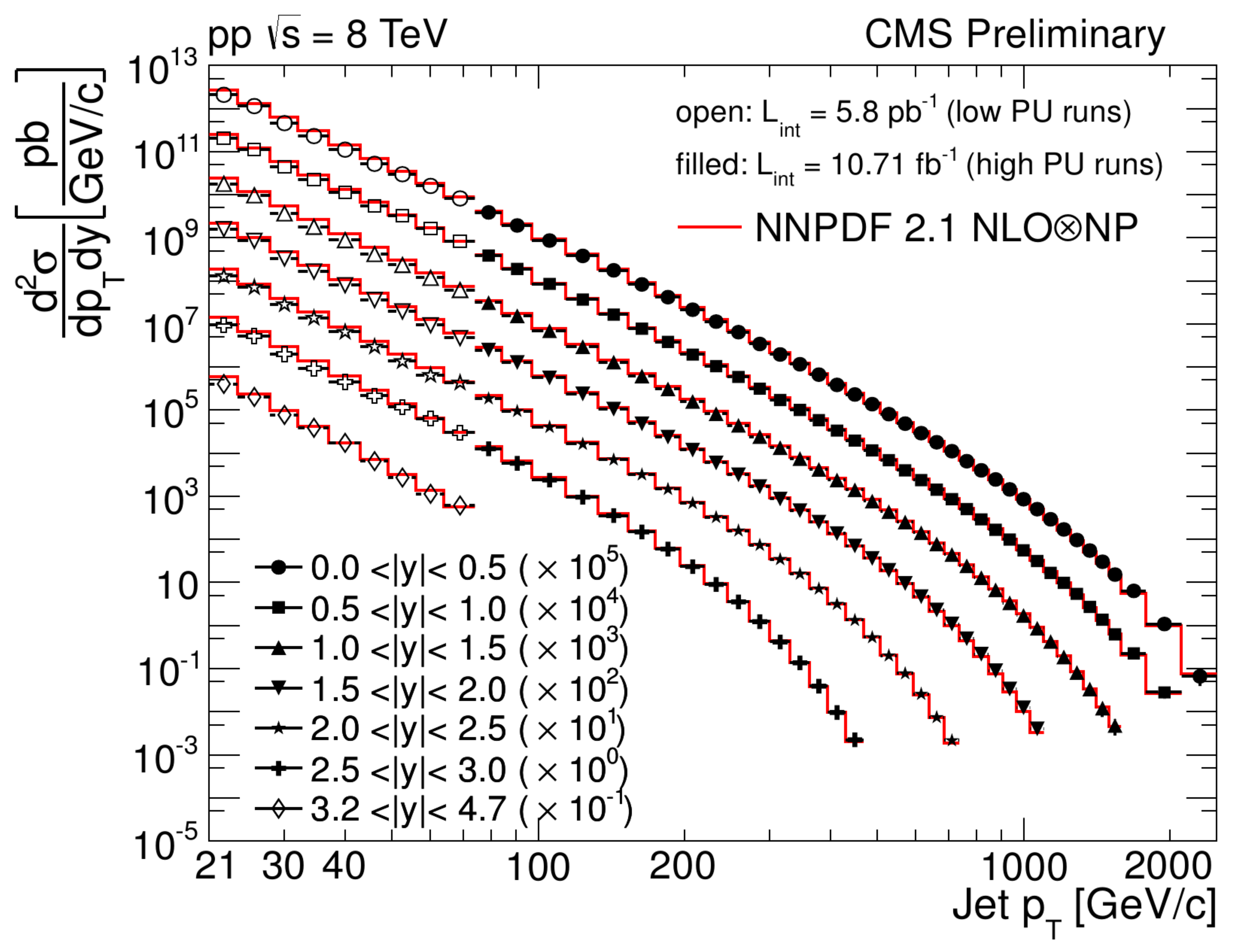} 
     \caption{Ratio of inclusive jet cross section measurement to NLO calculation prediction with CT10 PDF set on the left. On the right inclusive jet cross-section measurement over the wide rapidity and $p_{\mathrm{T}}$ ranges is shown.}
    \label{fig:spec}
\end{figure}

The measurement of inclusive production of jets \cite{lowpt} was performed with $5.8$ pb$^{-1}$ of pp collisions at $\sqrt{s} = 8$ TeV, recorded in special low pileup LHC runs during the summer of 2012. The production cross section was measured in bins of transverse momentum, $p_{\mathrm{T}}$, in the rapidity range of $\vert y \vert < 4.7$.  
The total experimental uncertainty does not exceed 50\%, with the leading contribution being due to the jet energy scale calibration.
The results of the measurement for the very forward region of the detector are shown in Fig.~\ref{fig:spec}, in addition to a combination of the described measurement with the inclusive jet production cross section measured over the full 8 TeV dataset \cite{smpjets} is presented. The results of the measurement were compared to the predictions of NLO pQCD calculations corrected for non-perturbative effects. The theoretical predictions agree with the measurements within the wide range of $p_{\mathrm{T}}$ and rapidity.

\begin{figure}[hbtp]
   \includegraphics[width=0.3\textwidth]{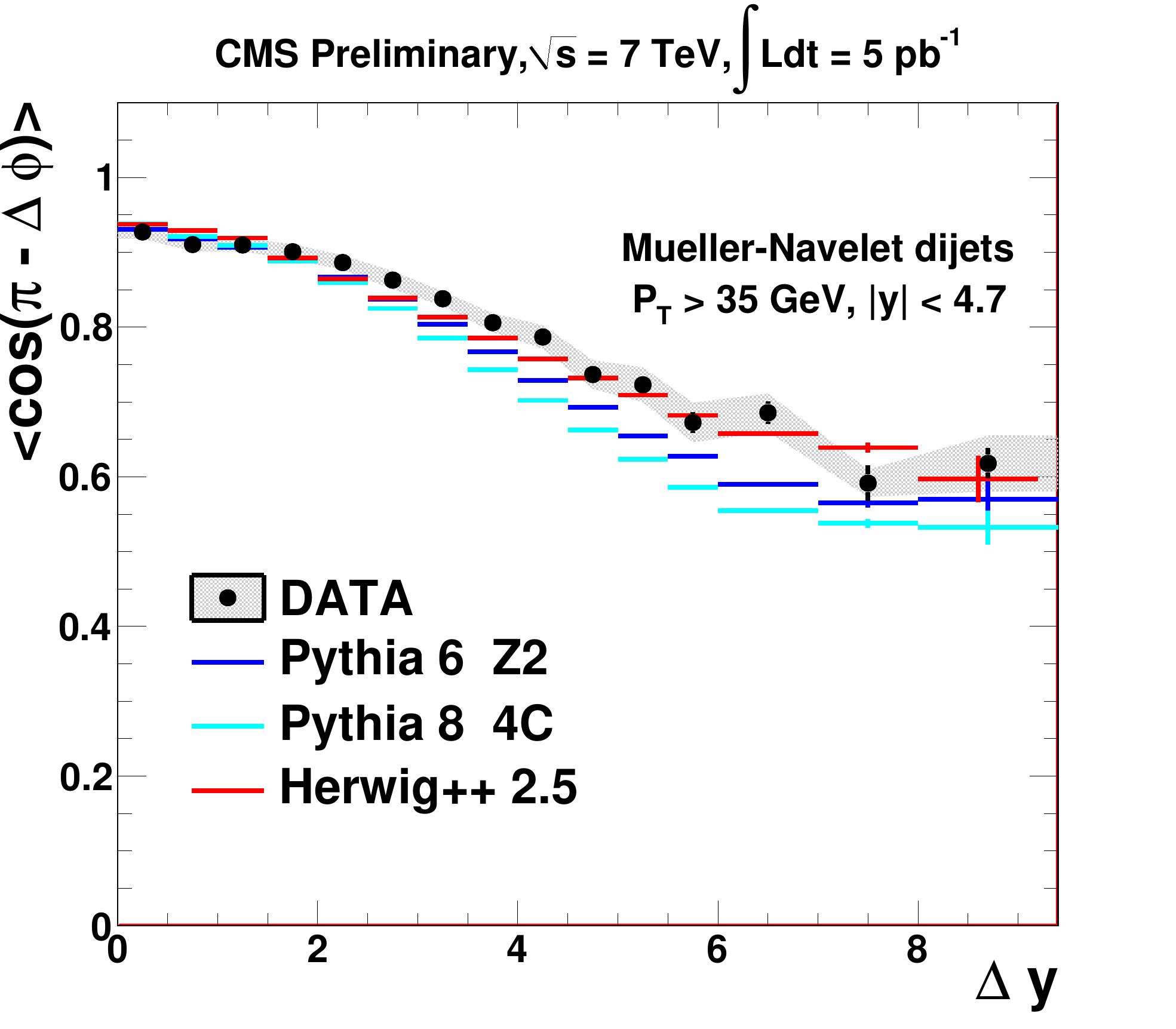}
    \includegraphics[width=0.3\textwidth]{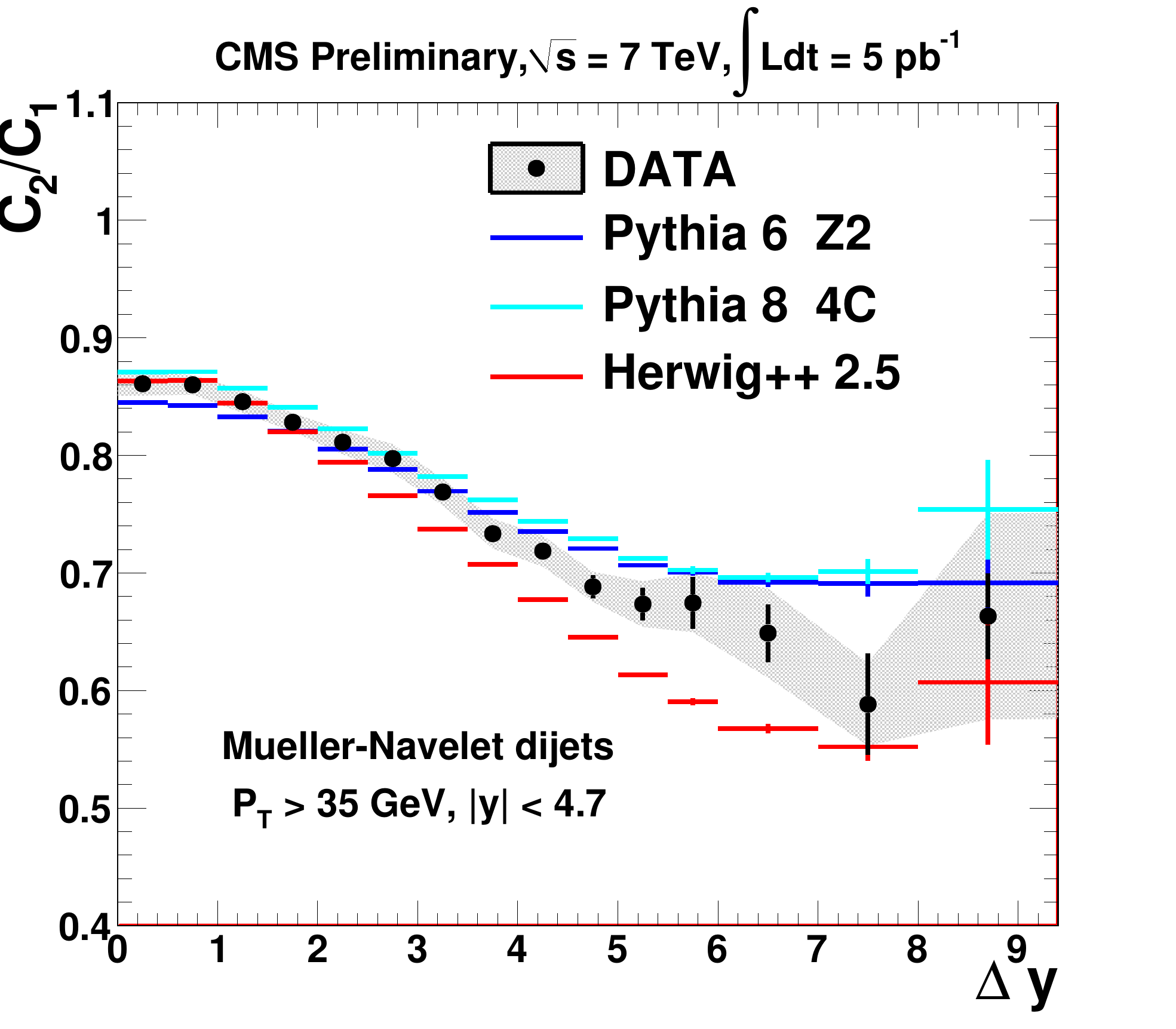} 
    \includegraphics[width=0.3\textwidth]{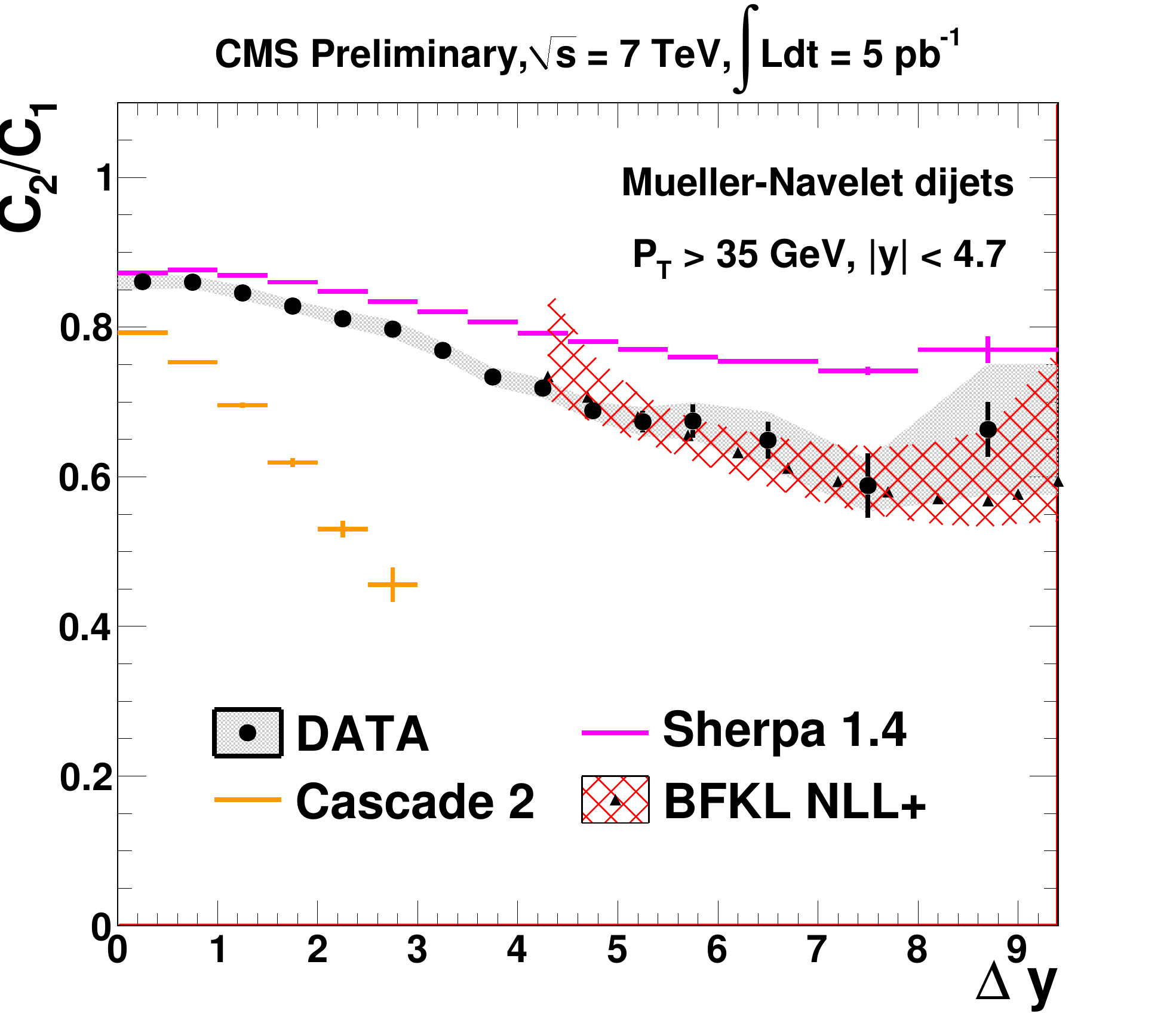} 
    \caption{$C_1$ as a function of $\Delta y$ (left). Ratio $C_2/C_1$ as a function of $\Delta y$ compared to various theory predictions (centre, right). }
    \label{fig:crat}
\end{figure}

\section{Decorrelation of Mueller-Navelet jets}
In high-energy pp collisions, a promising final state for probing BFKL asymptotic consists of two jets of similar $p_{\mathrm{T}}$ that are widely separated in rapidity, $y$, often referred to as Mueller-Navelet jets \cite{Mueller:1986ey}. In case of a significant contribution of semi-hard scattering in pp collisions at a given energy a strong rise of the inclusive dijet production cross section with increasing rapidity interval, $\Delta y$, is expected due to increased higher order parton radiation. Earlier CMS result probing higher order radiation in Mueller-Navelet jet production is inclusive to exclusive dijet production ratio measured as a function of rapidity separation between jets \cite{Chatrchyan:2012pb}. DGLAP-based \textsc{pythia} Monte Carlo predictions demonstrate good agreement with the data in that work. Another probe of higher order radiation is azimuthal decorrelation of jets within dijet pairs. Azimuthal decorrelation measured as a function of rapidity separation between jets may reveal effects beyond the DGLAP description.  

Average cosine of azimuthal separation between jets for particular $\Delta y$ range can show the level of angular decorrelation. The following quantities were measured as a function of $\Delta y$ in the work \cite{azimu}: $C_n = \langle \cos(n ( \pi - \Delta \phi))\rangle$, where $ \Delta \phi = \phi_1 - \phi_2 $ is the difference between the azimuthal angles $\phi_1$ and $\phi_2$ of the jets most forward and backward in rapidity and $n = 1, 2, 3$. A measurement of $C_1$ was first performed by the D0 experiment at the Tevatron \cite{Abachi:1996et}. Later it was proposed that the ratios of coefficients $C_2/C_1$ and $C_3/C_2$ are observables more sensitive to BFKL contributions~\cite{Vera:2007kn}. In this report a first measurement of cosine ratios is discussed, in addition to $C_n$. 
Analysis was performed with pp collisions taken at $\sqrt{s} = 7$ TeV in 2010. An effective integrated luminosity of 5 pb$^{-1}$ was available for the measurement. Average cosines of the azimuthal angle difference were measured in bins of of rapidity separation between the jets, $\Delta y$. Jets with $p_{\mathrm{T}}$ > 35 GeV and $\vert y \vert < 4.7$ were considered. 
Measured observables corrected for detector effects are compared to predictions of various MC and to analytic next-to-leading logarithmic approximation (NLL) BFKL calculations \cite{Ducloue:2013hia}. It should be noted that improved NLL BFKL calculations, including a comparison to the CMS data, were released after the present measurement was published \cite{Ducloue:2013bva}. 
The leading source of experimental uncertainty in the measurement is the jet energy scale calibration, which constitutes up to 24\% depending on observable and rapidity separation range. The total experimental uncertainty does not exceed 25\%.
The measured average cosines and cosine ratios are shown in Fig.~\ref{fig:crat}. 
DGLAP-based MC generators (\textsc{pythia6}, \textsc{herwig++} and \textsc{sherpa}) do not describe all observables within the experimental uncertainty. 
The NLL BFKL calculations describe well the cosine ratios but fail for the average cosines. It should be noted that the NLL BFKL calculations performed in \cite{Ducloue:2013bva} agree with the data for all observables. Full NLL BFKL calculation from another group of authors have become available recently \cite{ivanovpapa}. Somewhat worse agreement with the CMS data is demonstrated in latter work.  

\section{Conclusions}
Integrated low-$p_{\mathrm{T}}$ jet cross section probes transition region from perturbative to non-perturbative QCD domain. Predictions from Monte Carlo models show wide spread and the best description of the data is demonstrated by the cosmic ray MC \textsc{epos}. 
The experimental data on inclusive jet production in a wide range of $p_{\mathrm{T}}$ and rapidity is described by the pQCD NLO predictions within the experimental and theoretical uncertainties. Dijet production ratios and azimuthal decorrelations probe higher order pQCD radiation, with the theory predictions demonstrating a significant spread. 
The MC predictions show different levels of agreement with the data for different azimuthal decorrelation observables, with the best overall agreement demonstrated by \textsc{herwig++}. NLL BFKL calculations provide a good description of the average cosine ratios.

 \section{Acknowledgements}
 The author of these proceedings have received support from RFBR (Russia), grant number 14-02-31388.


\end{document}